%% file: main.tex
\documentclass{article}
\usepackage{spconf,amsmath,graphicx,booktabs,multirow,xcolor,url}
\usepackage{tikz}
\usetikzlibrary{arrows.meta,positioning}
\usepackage[hidelinks]{hyperref}



\newcommand{\nfn}{NFN}
\newcommand{\npc}{NPC}
\newcommand{\npca}{NPC-a}
\title{NORMALISE OR CONDITION? NOISE-FLOOR FRONT-ENDS FOR ON-BOARD KEYWORD SPOTTING UNDER UAV ROTOR EGO-NOISE}

\twoauthors{Yida Lin, Bing Xue, Mengjie Zhang}{Victoria University of Wellington\\ Wellington, New Zealand}{Sam Schofield, Richard Green}{University of Canterbury\\ Christchurch, New Zealand}

\begin{document}
\maketitle

\begin{abstract}
A microphone on the airframe of a small multi-rotor UAV is dominated by rotor ego-noise, so
spoken flight commands arrive at negative signal-to-noise ratio (SNR). We study small-footprint
keyword spotting (KWS) for a ten-word command vocabulary under \emph{real} ego-noise, training on
one quadrotor and testing on another. Besides per-clip accuracy we measure the streaming
false-alarm rate on 4.4\,h of continuous rotor noise. We compare classical noise-robust
front-ends (CMN, PCEN, spectral subtraction), test-time adaptation, and two front-ends that
track the per-band ego-noise floor over the two seconds preceding the decision window and
either subtract it (\emph{normalisation}) or feed it to the network as a second input channel
(\emph{conditioning}). Per clip, all front-ends look alike: $+$2--3 points on average, up to
$+$14 at $-15$\,dB. On continuous rotor noise they differ sharply. Normalising front-ends fire
about ten times more often than plain log-mel at the same threshold and end up \emph{below} it
at a budget of one false alarm per hour ($-$9 points at 0\,dB). Conditioning keeps the
baseline's false-alarm rate and turns its gain into detections ($+$8 points at $-10$\,dB over
three seeds); on top of PCEN it gives the best per-clip accuracy and false-alarm rate, and a
level-anchored variant is also invariant to the microphone gain. Real drone+interferer
recordings expose the remaining failure mode, environmental sounds and bystander speech, which
training negatives halve. A single script reproduces all on-device numbers on the NVIDIA Jetson
Orin NX flight computer, where the complete pipeline costs 3\,ms per 100\,ms hop on the CPU.
\end{abstract}


\section{Introduction}
\label{sec:intro}

Small multi-rotor UAVs are increasingly used for close-range vegetation work such as pruning
radiata pine, where on-board stereo vision detects and localises the branches to
cut~\cite{lin2025yolosgbm,lin2025segmentation}. The operator's hands and eyes are busy, so
short spoken commands (\emph{up}, \emph{down}, \emph{stop}, \dots) are a natural supplementary
control channel. Recognising them on a ground station adds a radio link, and its latency, to
the safety-critical \emph{stop}; we therefore spot the commands \emph{on board}, with one
microphone and the airframe's embedded computer (NVIDIA Jetson Orin).

The difficulty is the acoustic environment. Rotor ego-noise on a hovering multi-rotor is
broadband, harmonic, and 20--60\,dB louder than a person speaking a few metres away; in
DroneAudioset~\cite{gupta2025droneaudioset}, human sounds recorded by on-board microphones have
SNRs between $-57$ and $-2.5$\,dB. Classical drone audition attacks this with microphone
arrays and beamforming~\cite{hioka2016uav,wang2018drone,strauss2018dregon} or rotor-informed
enhancement networks~\cite{gulli2025rotor}. Both need multi-channel hardware or motor
telemetry and aim at intelligible speech rather than at spotting a handful of words.

Keyword spotting (KWS) is a closed-set problem on which very small networks excel on Google
Speech Commands (GSC)~\cite{warden2018speech}: broadcasted residual
networks~\cite{kim2021bcresnet}, keyword transformers~\cite{berg2021kwt} and state-space
models~\cite{ding2026kwm} reach 97--98\,\% on the 12-class task with 0.05--0.5\,M parameters.
Their weakness is \emph{unseen} noise. The classical answers are input normalisations---cepstral
mean normalisation (CMN)~\cite{viikki1998cmvn}, spectral subtraction~\cite{boll1979suppression}
with a minimum-statistics estimate~\cite{martin2001minstat}, per-channel energy normalisation
(PCEN) for far-field KWS~\cite{wang2017pcen}---and noise-aware training (NAT), which appends a
noise estimate to the input~\cite{seltzer2013nat}. Recent KWS work adds self-supervised
pre-training~\cite{mork2024noise}, noise-aware decoding~\cite{xi2025ntckws} and test-time
adaptation (TTA) of normalisation statistics~\cite{li2018adabn,xiao2025adakws,ding2026imkws}.
Voice-controlled drones, in turn, rely on cloud or laptop speech-to-text
pipelines~\cite{simoes2024voice,park2020uav,henry2026voice}. None of this work evaluates KWS
under the ego-noise of the microphone that actually flies.

This paper does not claim a new network. Its contributions are: (i)~a reproducible benchmark
for on-board UAV command spotting: GSC words under \emph{real} ego-noise of two quadrotors from
DroneAudioset~\cite{gupta2025droneaudioset}, trained on one drone and tested on the other at
$-15$ to $20$\,dB, scored by per-clip accuracy \emph{and} by a streaming false-alarms-per-hour
evaluation on 4.4\,h of rotor noise; (ii)~two noise-floor front-ends that revive minimum
statistics and NAT for this setting: a per-band floor tracked over the two seconds
\emph{preceding} the decision window is either subtracted in the log domain (normalisation,
\nfn) or fed as a second, level-anchored input channel (conditioning, \npc/\npca), and the same
floor is simulated during ego-noise multi-condition training; (iii)~a systematic comparison
against CMN, PCEN, spectral subtraction, FiLM conditioning, generic-noise/clean training and
AdaBN test-time adaptation, with ablations of floor statistic, microphone gain, stale floors,
real interferers and model size over three seeds; and (iv)~a streaming implementation on the
Jetson Orin NX flight computer with measured cost, whose on-device replay reproduces the PC
evaluation. Code, models and the on-device script will be released.

\section{On-board keyword spotting under ego-noise}
\label{sec:method}

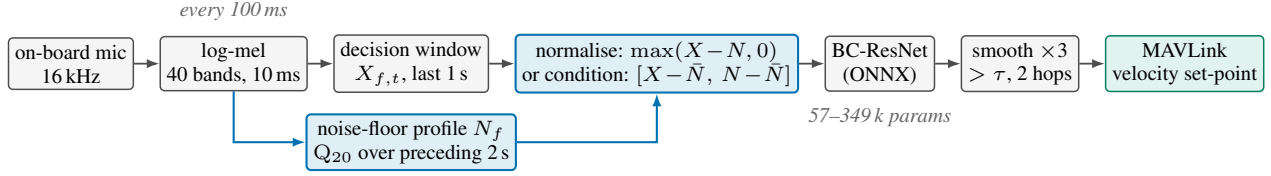
\begin{figure*}[t]
\centering
\resizebox{0.95\textwidth}{!}{\input{figs/pipeline.tex}}
\caption{On-board pipeline. Every 100\,ms the last second of log-mel frames is combined with
the noise-floor profile $N_f$ tracked over the preceding two seconds (blue blocks are the
proposed front-end) and classified; smoothed posteriors drive the flight controller over MAVLink.}
\label{fig:pipeline}
\end{figure*}

\subsection{Task and data}
\label{ssec:task}

The vocabulary is $\mathcal{V}=\{$\emph{up, down, left, right, forward, backward, go, stop, on,
off}$\}$ plus \emph{unknown} (any other GSC word) and \emph{silence} (ego-noise without
speech), i.e.\ 12 classes as in the standard GSC protocol. All words exist in
GSC~v2~\cite{warden2018speech}; we use the official train/validation/test lists
(32.4\,k/3.8\,k/4.3\,k clips after re-balancing \emph{unknown} and \emph{silence} to the mean
per-word count; the \emph{unknown} pool is re-sampled every epoch). Ego-noise comes from the
\emph{drone-only} subset of DroneAudioset~\cite{gupta2025droneaudioset}: 8.8\,h of 16\,kHz
recordings of two quadrotors at two microphone distances (25/50\,cm from the rotors), two
throttle levels and three microphones. Training and validation noise come from \emph{drone2}
(disjoint files); test noise comes exclusively from the unseen \emph{drone1}. Mixtures are
created on the fly. The noise is normalised to a fixed RMS, the level the airframe microphone
sees, and the speech is scaled so that its \emph{active} power (frames within 30\,dB of the
loudest) reaches the target SNR: uniform in $[-10,20]$\,dB plus 10\,\% clean clips for
training, $\{-15,\dots,20\}$\,dB and clean for testing. \emph{Silence} examples are noise-only
segments, so their error is the false-alarm rate (FAR) on rotor noise. Speech is added
synthetically (no Lombard effect or propeller wash on the talker); Sec.~\ref{ssec:beyond} adds
real drone+loudspeaker recordings.

\subsection{Front-end: tracking and using the ego-noise floor}
\label{ssec:frontend}

Let $X_{f,t}$ (dB) be the 40-band log-mel spectrogram (30\,ms window, 10\,ms hop, 98 frames per
decision window). Ego-noise changes slowly compared with a command, so its per-band level can
be estimated from the audio that \emph{precedes} the window, during which nobody is (usually)
speaking. Following the minimum-statistics idea~\cite{martin2001minstat} we define the
noise-floor profile
\begin{equation}
  N_f = \mathrm{Q}_{20}\big(\{X_{f,\tau}\}_{\tau\in\mathcal{C}}\big),
  \label{eq:profile}
\end{equation}
the 20th percentile over the $|\mathcal{C}|=200$ frames (2\,s) before the window. A low
percentile is insensitive to sporadic speech inside $\mathcal{C}$; during training a random GSC
clip is inserted into the context with probability 0.3. Three ways of using $N_f$ are compared, with fixed constants
$(\mu,\sigma)=(-20,25)$\,dB:
\begin{itemize}\setlength\itemsep{0pt}
\item \textbf{\nfn{} (noise-floor normalisation)}: input $\max(X_{f,t}-N_f,0)/\sigma$, i.e.\
  $10\log_{10}(1+\mathrm{SNR}_{f,t})$ up to the estimation error. This is log-domain spectral
  subtraction~\cite{boll1979suppression}, invariant to the noise level and microphone gain.
\item \textbf{\npc{} (noise-profile conditioning)}: two input channels
  $[(X_{f,t}-\mu)/\sigma,\ (N_f-\mu)/\sigma]$, $N_f$ broadcast over time. The first convolution
  sees the floor next to the noisy image and learns its own compensation, but it still sees
  absolute levels.
\item \textbf{\npca{} (level-anchored \npc)}: both channels relative to the overall floor level
  $\bar N=\frac1F\sum_f N_f$, i.e.\ $[(X_{f,t}-\bar N)/\sigma,\ (N_f-\bar N)/\sigma]$. A gain
  change shifts $X$, $N_f$ and $\bar N$ alike and cancels, so \npca{} combines the invariance of
  \nfn{} with the conditioning of \npc.
\end{itemize}
The network is unchanged except for the number of input channels ($<$1\,k extra parameters);
the baseline (\emph{log-mel}) uses $(X_{f,t}-\mu)/\sigma$.

\textbf{Relation to prior art and compared front-ends.} \nfn{} is spectral subtraction moved
to the log-mel domain with a minimum-statistics-like estimate, and \npc{} is noise-aware
training~\cite{seltzer2013nat}. What is new is where the estimate comes from---the two seconds
\emph{before} the window---and that the same statistic is simulated during multi-condition
training, with speech in the context, so that the network learns its bias. We compare, under
identical training, against \emph{CMN}~\cite{viikki1998cmvn} ($X_{f,t}-\bar X_f$ over the
window, also gain-invariant), \emph{PCEN}~\cite{wang2017pcen} (fixed $\alpha{=}0.98$,
$\delta{=}2$, $r{=}0.5$, $s{=}0.025$, smoother initialised on the 2\,s context), linear-domain
\emph{spectral subtraction} $10\log_{10}\max(P-N,0.01P)$ with the same
floor~\cite{boll1979suppression}, FiLM conditioning~\cite{perez2018film} (the floor scales and
shifts the first feature maps through a small MLP), and AdaBN~\cite{li2018adabn} test-time
adaptation, i.e.\ BatchNorm statistics re-estimated on unlabeled target-drone audio (the first
step of AdaKWS~\cite{xiao2025adakws}).

\subsection{Networks and training}
\label{ssec:nets}

We use three small-footprint back-ends from the public Keyword-Mamba code
base~\cite{ding2026kwm}: BC-ResNet-3/-8~\cite{kim2021bcresnet} (57\,k / 349\,k parameters) and
KWT-1~\cite{berg2021kwt} (0.56\,M), trained from scratch for 50 epochs with AdamW (lr
$10^{-3}$, weight decay 0.1, 5 warm-up epochs, cosine decay, batch 256), label smoothing 0.1,
$\pm$100\,ms time shift, $\pm$6\,dB gain and SpecAugment~\cite{park2019specaugment}, keeping
the checkpoint with the best validation accuracy over $\{-10,\dots,10\}$\,dB and clean.

\subsection{Streaming decision logic on the UAV}
\label{ssec:stream}

On the Jetson the microphone is read in 100\,ms blocks; log-mel frames, the running profile
(\ref{eq:profile}) and the network input are computed incrementally in NumPy and the ONNX model
is evaluated every hop. Posteriors are averaged over three hops~\cite{chen2014kws}; a command
fires when the smoothed posterior of a command class exceeds a threshold $\tau$ (0.7 by default)
in two consecutive hops, after which the same word is inhibited for 1\,s. Commands become
body-frame MAVLink velocity set-points; \emph{stop} is always accepted, \emph{on}/\emph{off} gate
motion.

\section{Experiments}
\label{sec:exp}
\begin{table*}[t]
\centering
\caption{BC-ResNet-8 on the 12-class UAV command task, test words mixed with \emph{unseen-drone} ego-noise:
accuracy (\%) per SNR, mean over the seven noisy SNRs, and false-alarm rate FAR (\%) on
noise-only input. Train noise: E = drone2 ego-noise, G = GSC noise, -- = clean; $\pm$ = std
over three seeds; bold = best in the E block.}
\label{tab:frontend}
\footnotesize
\setlength{\tabcolsep}{5pt}
\begin{tabular}{llcccccccc|cc}
\toprule
Front-end & Train & $-15$ & $-10$ & $-5$ & 0 & 5 & 10 & 20 & clean & mean & FAR \\
\midrule
\input{tables/frontend.tex} 
\bottomrule
\end{tabular}
\end{table*}

\begin{figure}[t]
\centering
\includegraphics[width=\columnwidth]{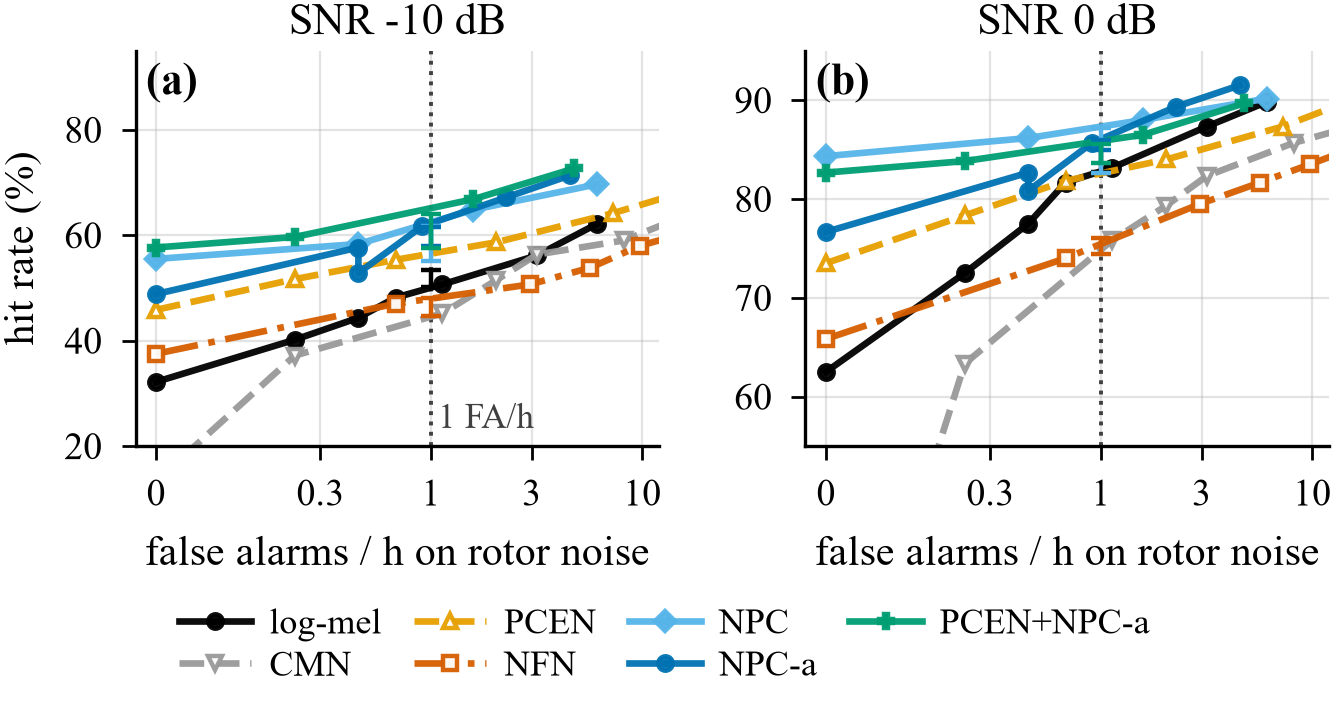}
\caption{Streaming operation (BC-ResNet-8, deployed decision logic): hit rate of commands
embedded every 3\,s in unseen-drone rotor noise vs.\ false alarms per hour on 4.4\,h of
speech-free rotor noise; one marker per threshold $\tau\in[0.5,0.95]$ (``0'': none in 4.4\,h),
curves for seed 0, error bars: std over three seeds at 1\,FA/h. Open markers: normalising.}
\label{fig:streaming}
\end{figure}

\begin{figure}[t]
\centering
\includegraphics[width=\columnwidth]{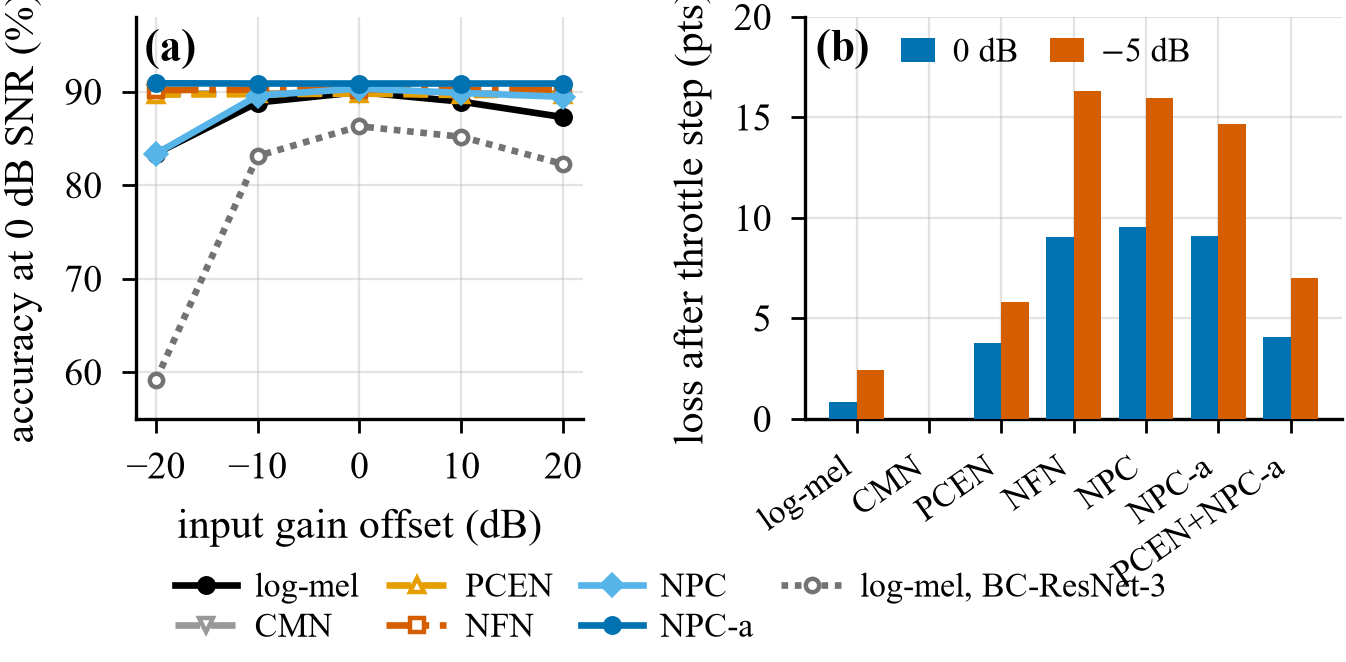}
\caption{Stress tests (BC-ResNet-8 unless noted). (a) Accuracy at 0\,dB SNR with the test
input scaled by $-20$ to $+20$\,dB. (b) Accuracy lost when the floor context comes from the
other throttle level of the same drone (mean over both directions).}
\label{fig:stress}
\end{figure}

\begin{table}[t]
\centering
\caption{Beyond rotor noise (BC-ResNet-8 unless noted, unseen drone, $\tau{=}0.7$): per-clip
mean accuracy; streaming hit rate at 0\,dB under 1\,FA/h; false alarms per hour on rotor noise,
on \emph{real} drone+non-speech-interferer and drone+bystander-speech recordings; firings per
100 non-command words. ``+ neg.'': ESC-50 events and overlapping non-command words as negatives.}
\label{tab:negatives}
\footnotesize
\setlength{\tabcolsep}{2.6pt}
\begin{tabular}{lcc|ccc|c}
\toprule
 & acc. & hit$_{0\,\mathrm{dB}}$ & \multicolumn{3}{c|}{FA/h @0.7} & unk. \\
Front-end & mean & 1\,FA/h & rotor & non-sp. & speech & /100 \\
\midrule
\input{tables/negatives.tex} 
\bottomrule
\end{tabular}
\end{table}

\subsection{Per-clip accuracy: training noise and front-ends}
\label{ssec:ablation}

Table~\ref{tab:frontend} reports BC-ResNet-8 on the test words mixed with the \emph{unseen}
drone's ego-noise. \emph{Training noise matters most.} A model trained on clean speech reaches
98.0\,\% on clean words, as expected from the literature~\cite{kim2021bcresnet}, but collapses
under rotor noise (61\,\% at 0\,dB, 16\,\% at $-15$\,dB) and calls \emph{every} noise-only
segment a word (FAR 100\,\%). Generic GSC noise (G) recovers part of the loss (72.3\,\%, FAR
43\,\%). The \emph{other} drone's ego-noise (E) is far better (81.2\,\%, FAR 3.6\,\%) although
the test drone was never seen. \emph{Under E, nearly all noise-aware front-ends look alike.}
They add $+$1.9 to $+$2.8 points on average (\nfn{} 83.2, \npc{} 83.1, \npca{} 83.5, CMN 83.5,
PCEN 84.0\,\%), mostly below 0\,dB ($+$14 at $-15$\,dB for PCEN, $+$9 for \npca), and stay
within $\pm$1 point above 5\,dB. Two exceptions: linear-domain spectral subtraction with the
same floor is worse than the baseline (79.1\,\%), so the non-negative log-domain form of \nfn{}
matters, and PCEN's compression costs 1.5 points on clean speech. \emph{The false-alarm rate
separates the families.} CMN and \nfn{} raise the baseline's FAR (3.6\,$\pm$\,2.1\,\%) to
7.3/6.9\,\%; conditioning keeps or lowers it (\npc{} 4.9, \npca{} 2.8\,\%). Subtracting the
floor removes the level evidence that separates a noise-only window from a faint word, whereas
the same floor given as a reference channel makes the ``no speech'' decision easier. With G
as the only training noise, \nfn{}/\npc{} still gain (77.5/78.1 vs.\ 72.3\,\%); on a
clean-trained model, \nfn{} \emph{hurts} (40.2 vs.\ 54.0\,\%). No front-end replaces
multi-condition training.

\textbf{Test-time adaptation.} AdaBN on unlabeled target-drone audio changes the
ego-noise-trained models by $-$0.4 to $-$1.1 points but helps the G-trained model by $+$3.3
(72.4 $\rightarrow$ 75.7\,\%), the unseen-noise-\emph{type} case TTA
targets~\cite{xiao2025adakws,ding2026imkws}.

\textbf{Floor statistic and its limits.} For \nfn, the 5th percentile under-estimates the floor
(81.8\,\%, FAR 10.4\,\%). The median (83.8\,\%) and even the plain context mean (84.2\,\%, FAR
6.2\,\%) beat the 20th-percentile default: once the statistic is simulated in training, the
minimum-statistics argument is moot and the mean is the cheapest tracker (\npca{} with the
mean: 84.2\,\%, FAR 0.6\,\%). A 0.5\,s context gives 82.7\,\%. Training without speech in the
context keeps accuracy (82.8\,\%) but raises the FAR to 9.8\,\%. \emph{Stale floor}
(Fig.~\ref{fig:stress}b): if the throttle changes right before a command, the floor-based
front-ends lose 9--16 points at 0/$-$5\,dB until the tracker catches up; log-mel and CMN are
unaffected and PCEN loses 4--6. A 2\,s
memory buys a 2\,s blind spot after every throttle step; a reset on level jumps is the remedy.

\textbf{Microphone-gain mismatch} (Fig.~\ref{fig:stress}a). Scaling the test input by
$\pm$20\,dB costs plain log-mel, trained with $\pm$6\,dB gain augmentation, up to 6.6 points at
0\,dB SNR, and \npc{} inherits this. CMN, PCEN, \nfn{} and \npca{} are invariant. For
BC-ResNet-3 the log-mel drop grows to 27 points (86.3 $\rightarrow$ 59.2\,\%) versus a flat
85.5/87.3\,\% with \nfn/\npca.

\subsection{Streaming operation: hit rate at a fixed false-alarm budget}
\label{ssec:streaming}

Per-clip accuracy hides the cost of false alarms, which on a UAV means unwanted manoeuvres.
Fig.~\ref{fig:streaming} therefore runs the complete streaming detector of
Sec.~\ref{ssec:stream} on continuous audio and plots, for every threshold, the false alarms per
hour on 4.4\,h of speech-free unseen-drone noise against the hit rate of commands embedded
every 3\,s into it (whole curves, so no operating point is tuned on test data). The picture
changes. At threshold 0.7, plain log-mel fires 0.8 times per hour on rotor noise and
\npc/\npca{} 1.0 times, but \nfn{} 7.7 and CMN 8.2 times. At a budget of 1\,FA/h the
normalising front-ends therefore fall \emph{below} plain log-mel (0\,dB: \nfn{} 75.3, CMN 74.9
vs.\ 84.2\,\%): the higher threshold they need costs more hits than the front-end gained.
Conditioning keeps the baseline's false-alarm rate and converts its per-clip gain into
detections. \npca{} reaches 59.7\,\% at $-$10\,dB ($+$8.0 over log-mel, $+$13 over \nfn) and
85.5/91.1\,\% at 0/5\,dB ($+$1.3, within the seed spread), with the smallest seed variance;
\npc{} without level anchoring is similar but noisier. PCEN, the best front-end per clip, lands
between the families (56.5/82.6\,\% at $-$10/0\,dB). \emph{Conditioning composes with normalisation.} Adding the anchored floor
channel to PCEN (PCEN+\npca) gives the best per-clip result, 84.5\,$\pm$\,0.2\,\% with FAR
1.5\,$\pm$\,0.1\,\%, and 60.8\,$\pm$\,3.2 / 84.6\,$\pm$\,0.9\,\% at $-$10/0\,dB under 1\,FA/h
($+$9 over log-mel at $-$10\,dB, on par with \npca); it also halves the stale-floor loss
(Fig.~\ref{fig:stress}b) because PCEN's own smoother adapts within the window. \emph{How} the
floor is injected matters too. FiLM conditioning (per-channel scale/shift of the first feature
maps) improves neither per-clip accuracy (81.0\,\%) nor the streaming gain (55.7\,\% at
$-$10\,dB), because it cannot express a per-band compensation. We therefore recommend
conditioning on the floor as a band-aligned input, on top of a normalisation.

\subsection{Different back-ends}
\label{ssec:backends}

For the 57\,k BC-ResNet-3 the mean accuracy is 75.9\,\% with log-mel, 78.9 with \nfn, 77.7 with \npc{} and 79.2 with \npca{} (FAR 2.5/5.0/5.0/2.0\,\%). In streaming mode \nfn{} again pays for its gain with false alarms
(72.1\,\% at 0\,dB under 1\,FA/h vs.\ 83.3 for log-mel), whereas \npca{} reaches 55.0/85.4\,\%
at $-$10/0\,dB ($+$8.0/$+$2.1) and thus matches the six-times larger BC-ResNet-8 baseline. KWT-1 (558\,k) reaches only 74.2/75.1\,\% (log-mel/\npca) even after 120 epochs.

\subsection{Beyond rotor noise: real recordings and negatives}
\label{ssec:beyond}

DroneAudioset also contains recordings in which a loudspeaker plays
Librispeech~\cite{panayotov2015librispeech} speech, crying, human non-vocal sounds (knocks,
footsteps) and non-human sounds (alarms, traffic) while the drone runs: real acoustic mixtures
with segment annotations. We replay the 2.0\,h of \emph{drone1} recordings through the
streaming detector and count firings inside non-speech segments (unambiguous false alarms) and
inside speech segments (bystander speech; it contains command words, so this rate is only
comparable across systems). Table~\ref{tab:negatives} shows the failure mode that
pure-noise testing misses. Every front-end fires 25--50 times per hour on real environmental
sounds and $>$100 times per hour on nearby speech, although only about once per hour on rotor
noise alone: what a model has never seen as a negative it cannot reject, whatever the
front-end. Adding negatives to training---ESC-50~\cite{piczak2015esc} sound events mixed into
the window (label unchanged) and a second, overlapping non-command word on \emph{unknown}
samples---roughly halves these rates (\npca: 42 $\rightarrow$ 27 non-speech and 132
$\rightarrow$ 98 speech triggers per hour; BC-ResNet-3 \npca: 29 $\rightarrow$ 13 and 119
$\rightarrow$ 51) at an unchanged rotor-noise false-alarm rate, for 0.5--0.7 points per clip.
Non-command \emph{words} still fire 4--6 times per 100 utterances for every variant; together
with the remaining bystander-speech triggers, this argues for a wake word rather than for a
better front-end.

\subsection{Deployment on the Jetson Orin NX}
\label{ssec:jetson}

The networks are exported to ONNX. The NumPy front-end reproduces the training features to
$10^{-5}$\,dB, and a single script on the flight computer measures the latency of every model
per execution provider and replays a bundled test set (5\,min of unseen rotor noise,
3$\times$100 commands at $+$5/0/$-$5\,dB, real drone+interferer excerpts) through the actual
runtime. Table~\ref{tab:jetson} gives the per-hop cost on an NVIDIA Jetson Orin NX 16\,GB
(JetPack 6.2.1, MAXN, ONNX Runtime 1.24). On two CPU cores the complete BC-ResNet-3 pipeline
costs 3.2\,ms per 100\,ms hop including the front-end (3\,\% of the hop), leaving the GPU to
the branch-detection stack~\cite{lin2025yolosgbm}; BC-ResNet-8 costs 8.6\,ms. TensorRT (FP16) brings every network to 1.2--1.5\,ms (p90 up to 5\,ms); the plain
CUDA provider is \emph{slower} than the CPU (3.4--6.7\,ms) because launch overhead dominates
for networks this small, and INT8 quantisation brings nothing. The
on-device replay yields decisions identical to the PC evaluation, so
Table~\ref{tab:frontend}, Fig.~\ref{fig:streaming} and Table~\ref{tab:negatives} describe the
deployed system. End-to-end latency is set by the decision logic (two agreeing hops,
$\approx$200\,ms), not by the network.

\begin{table}[t]
\centering
\caption{Per-hop (100\,ms) processing time in ms, batch 1, median over 300--500 runs. PC: Intel
i5-14600K, ONNX Runtime CPU, 2 threads. Jetson: Orin NX 16\,GB, JetPack 6.2.1, MAXN; ONNX
Runtime 1.24 CPU (2 threads) or TensorRT execution provider (FP16). The front-end always runs
in NumPy on the CPU.}
\label{tab:jetson}
\small
\setlength{\tabcolsep}{2.5pt}
\begin{tabular}{lrrrr}
\toprule
Stage / model & Params & PC & Jetson & Jetson \\
 & & CPU & CPU & TensorRT \\
\midrule
Front-end (log-mel $+$ $N_f$) & -- & 0.10 & 0.57 & n/a \\
BC-ResNet-3 + \npca & 57\,k & 1.07 & 2.64 & 1.32 \\
BC-ResNet-8 + \npca & 349\,k & 2.09 & 8.01 & 1.45 \\
BC-ResNet-8 + PCEN+\npca & 349\,k & 2.02 & 8.10 & 1.44 \\
KWT-1 + \nfn & 558\,k & 1.56 & 6.61 & 0.75 \\
\bottomrule
\end{tabular}
\end{table}

\section{Conclusion}
\label{sec:concl}

Spotting flight commands next to the rotors is feasible with a 57--350\,k parameter network
if it is trained on recorded ego-noise (a 98\,\%-accurate clean model fires on every second of
hovering) and the front-end is chosen with the false-alarm budget in mind. Per
clip, all robust front-ends look alike. On continuous rotor noise the normalising ones fire ten
times more often than log-mel and fall \emph{below} it at a fixed false-alarm rate, whereas
conditioning the network on the tracked floor keeps the baseline's rate and turns the gain into
detections ($+$8--9 points at $-10$\,dB under 1\,FA/h, with or without PCEN). Real recordings
show what remains: bystander speech and environmental sounds trigger the detector dozens of
times per hour, training negatives halve that, and a wake word should absorb the rest; on the
Jetson Orin NX the pipeline costs 3\,ms per 100\,ms hop.

\bibliographystyle{IEEEbib}
\bibliography{refs}

\end{document}

%% file: figs/pipeline.tex
\definecolor{propblue}{HTML}{0072B2}
\definecolor{propgreen}{HTML}{009E73}
\begin{tikzpicture}[
  font=\scriptsize,
  box/.style={draw=black!70, line width=0.5pt, rounded corners=1.5pt, minimum height=5.6mm, inner sep=2.2pt, align=center, fill=black!4},
  prop/.style={box, draw=propblue, fill=propblue!12, line width=0.7pt},
  ctl/.style={box, draw=propgreen!80!black, fill=propgreen!12},
  arr/.style={-{Latex[length=1.5mm, width=1.1mm]}, line width=0.6pt, draw=black!70},
  parr/.style={arr, draw=propblue, line width=0.7pt},
  node distance=2.4mm and 3.0mm]
  \node[box] (mic) {on-board mic\\16\,kHz};
  \node[box, right=of mic] (lm) {log-mel\\40 bands, 10\,ms};
  \node[box, right=of lm] (win) {decision window\\$X_{f,t}$, last 1\,s};
  \node[prop, below=of win] (prof) {noise-floor profile $N_f$\\$\mathrm{Q}_{20}$ over preceding 2\,s};
  \node[prop, right=of win] (fe) {normalise: $\max(X\!-\!N,0)$\\or condition: $[X\!-\!\bar N,\,N\!-\!\bar N]$};
  \node[box, right=of fe] (net) {BC-ResNet\\(ONNX)};
  \node[box, right=of net] (dec) {smooth $\times$3\\$>\tau$, 2 hops};
  \node[ctl, right=of dec] (mav) {MAVLink\\velocity set-point};
  \draw[arr] (mic) -- (lm);
  \draw[arr] (lm) -- (win);
  \draw[parr] (lm.south) |- (prof.west);
  \draw[arr] (win) -- (fe);
  \draw[parr] (prof.east) -| (fe.south);
  \draw[arr] (fe) -- (net);
  \draw[arr] (net) -- (dec);
  \draw[arr] (dec) -- (mav);
  \node[above=0.6mm of lm, font=\scriptsize\itshape, text=black!60] {every 100\,ms};
  \node[below=0.6mm of net, font=\scriptsize\itshape, text=black!60] {57--349\,k params};
\end{tikzpicture}

%% file: tables/frontend.tex
log-mel & -- & 15.8 & 23.3 & 38.9 & 61.2 & 73.2 & 80.0 & 85.6 & 98.0 & 54.0 & 100.0 \\
\midrule
log-mel & G & 29.9 & 53.3 & 71.6 & 82.6 & 87.2 & 89.8 & 91.4 & 96.0 & 72.3 & 43.1 \\
\nfn & G & 44.5 & 65.7 & 78.3 & 85.2 & 88.2 & 89.6 & 90.7 & 96.6 & 77.5 & 53.8 \\
\npc & G & 41.8 & 64.2 & 78.8 & 86.8 & 90.3 & 91.9 & 93.1 & 96.3 & 78.1 & 30.3 \\
\midrule
log-mel & E & 42.8 & 69.9 & 83.5 & 89.6 & 92.8 & 94.6 & 95.7 & 96.7 & 81.2\,$\pm$\,0.2 & 3.6\,$\pm$\,2.1 \\
CMN & E & 55.3 & 73.2 & 84.0 & 89.6 & 92.6 & 94.2 & 95.3 & \textbf{97.0} & 83.5 & 7.3 \\
PCEN & E & 57.2 & 75.2 & 84.8 & 90.2 & 92.4 & 93.7 & 94.8 & 95.2 & 84.0 & 3.9 \\
spec.\ sub. & E & 37.0 & 66.9 & 81.1 & 88.6 & 91.9 & 93.3 & 94.9 & 96.4 & 79.1 & 8.7 \\
\nfn & E & 52.3 & 73.1 & 84.5 & 89.7 & 92.5 & 94.1 & 95.0 & 96.7 & 83.2\,$\pm$\,0.3 & 6.9\,$\pm$\,0.7 \\
\npc & E & 50.7 & 73.6 & 85.2 & 90.2 & 93.0 & 94.6 & 95.6 & 96.7 & 83.1\,$\pm$\,0.1 & 4.9\,$\pm$\,1.7 \\
\npca & E & 51.7 & 74.3 & \textbf{85.8} & \textbf{91.0} & \textbf{93.9} & \textbf{95.1} & \textbf{96.0} & 96.6 & 83.5\,$\pm$\,0.4 & 2.8\,$\pm$\,1.0 \\
PCEN+\npca & E & \textbf{59.1} & \textbf{76.9} & 85.7 & 89.9 & 92.4 & 93.7 & 95.2 & 95.4 & \textbf{84.5\,$\pm$\,0.2} & 1.5\,$\pm$\,0.1 \\%

%% file: tables/negatives.tex
log-mel & 81.3 & 82.8 & 1.1 & 27 & 143 & 3.8 \\
log-mel + neg. & 81.6 & 88.7 & 0.0 & 15 & 66 & 6.2 \\
\nfn & 83.0 & 75.4 & 9.8 & 40 & 124 & 5.2 \\
\nfn{} + neg. & 82.8 & 60.7 & 12.5 & 24 & 93 & 6.2 \\
\npca & 84.0 & 86.1 & 0.9 & 42 & 132 & 4.5 \\
\npca{} + neg. & 83.3 & 85.2 & 1.1 & 27 & 98 & 5.0 \\
PCEN+\npca & 84.7 & 85.9 & 0.2 & 36 & 123 & 5.0 \\
BC-ResNet-3 \npca & 79.2 & 85.4 & 0.2 & 29 & 119 & 6.2 \\
BC-ResNet-3 \npca{} + neg. & 78.7 & 78.2 & 0.5 & 13 & 51 & 7.2 \\%